\documentclass[conference]{IEEEtran}
\IEEEoverridecommandlockouts

\usepackage{amssymb}
\usepackage[cmex10]{amsmath}
\usepackage{stfloats}
\usepackage{graphicx}
\usepackage{subfigure}
\usepackage{tabularx}
\usepackage{epsfig,epsf,color,balance,cite}
\usepackage{verbatim}
\usepackage{url}
\usepackage{bm}

\newtheorem{theorem}{\bf Theorem}
\newtheorem{proposition}{\bf Proposition}
\newtheorem{lemma}{\bf Lemma}

\usepackage{algorithm}
\usepackage{algorithmic}

\usepackage{color}
\definecolor{myc1}{rgb}{0,0,0}

\begin{document}

\title{{Machine Learning for Predictive Deployment of UAVs with  Multiple Access}  }

\author{
\IEEEauthorblockN{Linyan Lu\IEEEauthorrefmark{1},
Zhaohui Yang\IEEEauthorrefmark{1},
Mingzhe Chen\IEEEauthorrefmark{2},
Zelin Zang\IEEEauthorrefmark{3},
and Mohammad Shikh-Bahaei\IEEEauthorrefmark{1}
                  }
\IEEEauthorblockA{\IEEEauthorrefmark{1}Centre for Telecommunications Research, Department of Engineering, King's College London, London WC2B 4BG, UK.}
\IEEEauthorblockA{\IEEEauthorrefmark{2}Chinese University of Hong Kong, Shenzhen, 518172, China,\\ and also with the Department of Electrical Engineering, Princeton University, Princeton, NJ, 08544, USA.}
\IEEEauthorblockA{\IEEEauthorrefmark{3}College of Computer Science and Technology, Zhejiang University of Technology, Hangzhou 310027, China.}
\IEEEauthorblockA{Emails:
lulinyan-yy@foxmail.com, yang.zhaohui@kcl.ac.uk,  mingzhec@princeton.edu, zangzelin@gmail.com, m.sbahaei@kcl.ac.uk}
\vspace{-2em}
}

\maketitle

\begin{abstract}
In this paper, a machine learning based deployment framework of unmanned aerial vehicles (UAVs) is studied. In the considered model, UAVs are deployed as flying base stations (BS) to offload heavy traffic from ground BSs. Due to time-varying traffic distribution, a long short-term memory (LSTM) based prediction algorithm is introduced to predict the future cellular traffic. To predict the user service distribution, a KEG algorithm, which is a joint K-means and expectation maximization (EM) algorithm based on Gaussian mixture model (GMM), is proposed for determining the service area of each UAV. Based on the predicted traffic, the optimal UAV positions are derived and three multi-access techniques are compared  so as to minimize the total transmit power. Simulation results show that the proposed method can reduce up to 24\% of the total power consumption compared to the conventional method without traffic prediction. Besides, rate splitting multiple access (RSMA) has the lower required transmit power compared to frequency domain multiple access (FDMA) and time domain multiple access (TDMA).
\end{abstract}

\IEEEpeerreviewmaketitle

\section{Introduction}
Since user demands for communication services grow dramatically,
traditional base stations (BSs) cannot meet the required demand of the cellular traffic, which can lead to a bottleneck of cellular communication \cite{saad2019vision,2,chen2020wireless}.
Recently, there has been a growing interest in the study of unmanned aerial vehicle (UAV) communication due to its excellent attributes of versatility, maneuverability and flexibility \cite{3}. UAV gradually plays an important role in wireless communications, which can offer low-cost and efficient wireless connectivity for devices. UAVs acting as flying BSs is one of the most important research objects in the UAV communication.
Through locations adjustment, obstacle avoidance, and line-of-sight (LoS) link reinforcement, UAVs are able to offload data traffic from loaded BSs
and increase the connectivity of wireless network  so as to improve the communication throughput, coverage, and energy efficiency \cite{20}.

Therefore,  it is a feasible and beneficial option to utilize UAVs to ensure the connectivity of wireless communication network via meeting the surging data demands.
For efficient and rapid dispatch of UAVs, the prediction of potential hotspot areas plays a crucial role to help network operators acquire the information of occurrence and degrees of congestion in advance to reduce the entire network communication delay \cite{4,8637952,8379427}.
Machine learning techniques is a useful tool which has the ability to efficiently predict the distribution of future traffic data \cite{8755300,dong2019deep,shi2020communication,jia2020channel,chen2019joint,yang2019energy,wang2019caching}.
With such predictions, the target locations of UAVs can be specified beforehand and the deployment can be more intelligent and on-demand.

There are a number of existing works investigating the applications of UAV deployment in communication. In UAV deployment, UAV service areas and their optimal placements are two critical factors \cite{8,7,wang2018energy,wang2019adaptive}. In \cite{8}, UAV latitudes and 2-D locations are optimized based on circle packing theory. Then, UAV placements and service area boundaries are separately determined for power efficiency in \cite{7}. To efficiently dispatch UAVs, some intelligent and practical methods are proposed. A functional split scheme selection of each UAV is jointed with UAV deployment in \cite{wang2018energy}. In \cite{wang2019adaptive}, an adaptive deployment scheme is introduced to optimize UAV displacement direction, time-dependent traffic of mobile users can be served in real time.

In UAV communication, machine learning techniques are also applied to improve system performance \cite{zhang2018predictive,wang2019deep,7875131}. The aerial channel environment is predom field position (CRF) in \cite{11}. In \cite{zhang2018predictive}, in order to provide more efficient downlink service, a learning approach based on the weighted expectation maximization (WEM) algorithm is used to predict downlink user traffic demand, meanwhile a traffic load contract using contract theory is introduced.

The main contribution of this work is a predictive UAV deployment scheme for UAV-assisted networking with ML approaches.
 Our key contributions include:
 \begin{itemize}
\item The cellular traffic is predicted according to the analysis of previous data by BP Neural Network model. Then, a joint K-means and EM algorithm of a Gaussian Mixture Model (KEG) is proposed to divide the entire service area into clusters as UAV service areas in the temporal and spatial patterns \cite{1}. UAV locations are optimized so as to minimize the total transmit power of the whole UAV swarm network. Three different access techniques have been compared in the simulation.
\item The results show that the proposed UAV deployment framework can reduce the overall transmit power by over 23.85\% compared to the conventional method. Besides, it is also shown that rate splitting multiple access (RSMA) can decrease up to 35.5\% and 66.4\% total power compared to frequency domain multiple access (FDMA) and time domain multiple access (TDMA), respectively.
\end{itemize}

\vspace{-.5em}
\section{System Model and Problem Formulation}
\vspace{-.25em}
Given a time-dependent UAV-assisted wireless network, a group of ground users in the network are distributed in a geographical area $C$. A set $\boldsymbol{I}$ of $I$ UAVs assist a set $\boldsymbol{J}$ of $J$ ground BSs to offload amounts of cellular traffic for congestion alleviation, so the ground users in a time-variant hotspot area have the air-ground communications with UAVs when ground BSs are overloaded.

It is assumed that the height of all users and BSs is zero compared to the UAVs. Besides, each UAV has directional antennas, so the transmissions of different UAVs will not be interfered with each other. For convenient elaboration, we specify the ground users served by a UAV as \emph{aerial users} and the service area of a UAV $i$ as an \emph{aerial cell} which can be expressed as $C_i$. In order to consider the communication of all users fairly, the aerial cells are supposed to completely cover the entire area without any overlaps.

Because UAVs have limited energy amounts, UAVs should be efficiently deployed

\subsection{Air-ground Model}
We assume that a ground user $j \in \boldsymbol{J}$ located at $(x,y)\in C$ and a UAV $i \in \boldsymbol{I}$ located at $(x_i,y_i,h_i)$ with the aerial cell $C_i$, so the uplink received power of a UAV $i$ and a ground user $j$ is calculated as:
\begin{equation} \label{power}
P_{r,ij} [dB] = P_{ij} [dB] + G_{ij}  [dB]  - L_{ij} [dB]  - r_{ij} [dB],
\end{equation}
where $P_{ij}$ is the transmit power, $G_{ij}$ is the antenna gain, $L_{ij}$ is the free space path loss, $r_i$ is the excessive path loss with a Gaussian distribution which depends on the category of link. For mathematical tractability, we give a hypothesis that all of beam alignment of ground-air links are perfect and UAV antenna gains are same. Thus, $G_{ij}$ can be a constant number $G$.
The free space path loss has a specific formula to be acquired:
\begin{equation} \label{free space path loss}
L_{ij}[dB] = 10n\log(\frac{4\pi f_c d_{ij}}{c}),
\end{equation}
where $n\ge2$ is the path loss exponent, $f_c$ is the system carrier frequency, $c$ is the light speed, $d_{ij} = [(x_i-x_j)^2 + (y_i-y_j)^2 + h_i^2)]^{-2}$ is the distance between a UAV $i$ and a user $j$ \cite{1}.

In general, the air-to-ground transmission is separated into two main cateogories: the line-of-sight link (LoS) and the non-line-of-sight link (NLoS) .
The NLoS link suffers a higher excessive path loss owing to shadowing and blockage. Figure 2 has shown these two links in the picture.
\begin{figure}[!htb]
	\centering\includegraphics[width=6cm]{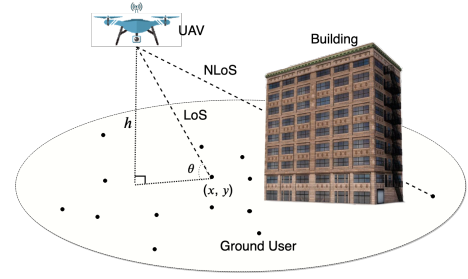}\\
	\caption{The LoS Link and the NLoS Link of UAV Transmission}
\end{figure}
The excessive path loss in these two links can be expressed as
${r_i^{LoS}} \sim {\cal N}(\mu_{LoS},\sigma_{LoS}^2)$ and ${r_i^{NLoS}} \sim {\cal N}(\mu_{NLoS},\sigma_{NLoS}^2)$ respectively.
The probability of the occurance of LoS link is similar to a sigmoid function \cite{36}:
\begin{equation} \label{probability}
p_{ij}^{LoS} = \frac{1}{1 + a \exp{[b (a - \theta_{ij})]}},
\end{equation}
where $a$ and $b$ are environment constant coefficients, $\theta_{e,ij} = sin^{-1}(h_i/d_{ij})$ is the elevation angle of a UAV $i$ and a user $j$, so the NLoS link existing probability is $p_{ij}^{NLoS} = 1 - p_{ij}^{LoS}$ \cite{1}. Therefore, the average excessive path loss in the uplink transmission is:
\begin{equation} \label{excessive path loss}
r_{ij} = {r_i^{LoS}}p_{ij}^{LoS} + {r_i^{NLoS}}p_{ij}^{NLoS}
\end{equation}

To this end, the uplink data rate of a UAV $i$ and a user $j$ can be expressed as:
\begin{equation} \label{data rate}
R_{ij} = W_{ij}\log_2 (\frac{P_{r,ij}}{N_0} + 1) (bits/s)
\end{equation}
where $W_{ij}$ is the bandwidth of a UAV $i$ and a user $j$ which can be , $N_0 = n_0W_{ij}$ is the power of addictive white Gaussian noise and $n_0$ is its average power spectral density.
For tractable formulation, each UAV can offer sufficient overall bandwidth and all transmit bandwidths are assumed to be a constant value $W$.

\subsection{Multi-access Modes}
In this paper, three different multi-access techniques will be used. They are rate splitting multiple access (RSMA), frequency division multiple access (FDMA) and time division multiple access (TDMA) \cite{45}. We set two users as an example to give a series of theoretical formulas and set $R_1$ and $R_2$ are the actual data rates of two users, $g_1$ and $g_2$ are the channel gains of two users which include the antenna gain, free space path loss and the excessive path loss.

RSMA is a multi-access mode with coding and decoding techniques, the equations for maximizing the sum-rate of users for uplink transmission are shown below \cite{yang2019optimization,45,7470942,mao2018rate,mao2019rate,8846706,8491100,8008852,clerckx2019rate}:
\begin{equation}
\begin{cases}
    R_1\leq W\log_2(1+g_1P_1/(Wn_0))\\
    R_2\leq W\log_2(1+g_2P_2/(Wn_0))\\
    R_1+R_2\leq W\log_2(1+(g_1P_1+g_2P_2)/(Wn_0))
\end{cases}
\end{equation}
FDMA is a technique of BS bandwidth allocation for users to avoid interference of transmissions in the same area, where $b_1$ and $b_2$ are the weight coefficients.
\begin{equation}
\begin{cases}
    R_1\leq Wb_1\log_2(1+g_1P_1/(Wn_0b_1))\\
    R_2\leq Wb_2\log_2(1+g_2P_2/(Wn_0b_2))\\
    b_1+b_2=1, b_1\geq0, b_2\geq0
\end{cases}
\end{equation}
Similarly, TDMA assigns the partial time period to users who can use the whole bandwidth, the equations have been represented.
\begin{equation}
\begin{cases}
    R_1\leq Wb_1\log_2(1+g_1P_1/(Wn_0))\\
    R_2\leq Wb_2\log_2(1+g_2P_2/(Wn_0))\\
    b_1+b_2=1, b_1\geq0, b_2\geq0
\end{cases}
\end{equation}

In simulation and analysis section, we will compare these three multi-access modes for choosing a best one to minimize the total transmit power.

\subsection{Uplink Transmit Power Computation}
Network operators require to assign UAVs to those traffic congestion areas so as to offload the heavy traffic from busy BSs. For UAVs, continuous movements will consume too much transmit power. Thus, we need to analyze the situation of traffic offloading. A data set is presented as a matrix $\boldsymbol{D}$:
\begin{equation}
\boldsymbol{D}= \{D^t_d(x,y) \mid t \in \{T,..., 24T\}, d \in \{1,..., 8\}, (x,y)\in C\}
\end{equation}
where $T$ is the time of an hour, $t$ is the time moment and $d$ is the day. $D^t_d(x,y)$ represents the amount of cellular traffic offloaded from a BS located at $(x,y)$ in a period of $T$ in the day $d$ \cite{1}. In this paper, for the conveinience of simulation and analysis, we assume that all cellular traffic of BSs are totally offloaded to UAVs.

Since the positions of mobile users are uncertain and most of mobile users only move around a single BS in a period of an hour, we assume that all ground receivers have the same positions as their nearst BSs.

After obtaining the future predictive traffic information through the implementation of ML methods based on the matrix $\boldsymbol{S}$, the required average data rate within a aerial cell $C_i$ in a period of $T$ will be given. Only when the communication throughput of UAV is not less than the demanded data rate, the communication can be ensured. Therefore, the communication condition is formulated as:
\begin{equation}
\iint_{C_i} R_i(x,y)dxdy \geq \frac{1}{T} \iint_{C_i} D^t_d(x,y)dxdy
\end{equation}
where R(x,y) is the maximum data rate of the overall transmission between a group of ground users in a ground cell with a BS at $(x,y)$ and a UAV $i$.
We can simplify this equation as
\begin{equation}
    R_i(x,y) \geq  D^t_d(x,y)/T
\end{equation}
Let us set $D^t_d(x,y)/T = \alpha (x,y)$, $\alpha (x,y)$ is the minimal requirement of average data rate. Combining the formula (3.5) and (3.10), the minimum power for transmission should be provided by a UAV
will be:
\begin{equation}
P_{i}^{min}(x,y) = (2^{\frac{\alpha (x,y)}W} - 1)n_0WL_{i}(x,y)r_i(x,y)/G
\end{equation}
where both $L_i(x,y)$ and $r_i(x,y)$ are between the UAV $i$ and the group of ground users in a ground cell of the BS at $(x,y)$, $L_i(x,y)$ is the path loss in free space, $r_i(x,y)$ is the excessive path loss. This equation provide a target basis for UAV location optimization.

This section proposes a novel predictive scheduling scheme of UAV based on ML.
According to the real data set in City Cellular Traffic Map \cite{chen2015analyzing} and the characteristics of celluar data traffic, for the sake of the efficiency of UAV deployment, we make some rational assumptions. Because humans have a certain pace of life with periodic acitivity, the change of cellular data traffic has a repetitive pattern in daily life \cite{44}. Thus, we assume that the cellular traffic amount has specific distribution in the same hour in different days and the data of each hour in the same day is independent. To this end, the real data set can be classified into 24 independent models and we assume each single model follows the Guassian mixture model which will be explained in section 4.3.

The logical procedure diagram of UAV predictive deployment is shown in Figure 3. At first, the acquired real data set is preprocessed to get the cellular traffic amount of every hour in the first 5 days $\{D^t_d\}_{d=1,t=T}^{5,24T}$ and the topology information of every BS $\{\boldsymbol{x}_n\}_{n=1}^N = \{(x_n,y_n)\}_{n=1}^N$, where $x_n$ is the relative longitude of $n^{th}$ BS and  $y_n$ is the relative latitude of $n^{th}$ BS. Then, a BP neural network model for cellular traffic amount prediction is come up to predict the cellular amount in $24$ hours in the $6^{th}$ day. At the same time, a joint K-means and EM algorithm relying on a GMM  for aerial cell classification (ground user clustering) is created, the point cluster label $\{l_n\}_{n=1}^N$ of every single point $\boldsymbol{x}_n$ is obtained, the point $\boldsymbol{x}_n$ with the same label value consists of an aerial cell. Then, the optimal UAV locations $\{\boldsymbol{x}_i\}_{i=1}^K=\{(x_i,y_i)\}_{i=1}^K$ for minimizing total transmit power $P_{min}$ is derived according to the system model introduced in section 3. The purpose of whole process is to achieve an UAV aid of the on-demand, power-effective, lantency-free network services.

\begin{figure}[!htb]
	\centering\includegraphics[width=6cm]{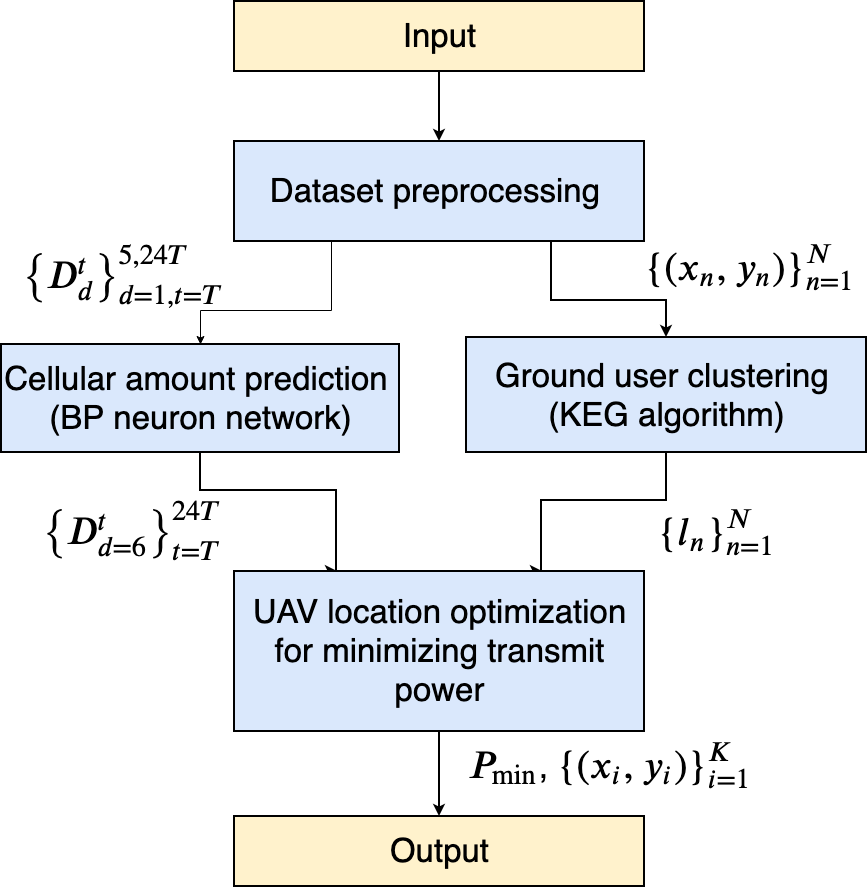}\\
	\caption{The Logical Procedure Diagram of UAV Predictive Deployment}
\end{figure}

\subsection{Cellular Demand Prediction}
In this part, a simple BP neural network model is utilized to predict the future cellular demand in an hour. Neural network has a variety of categories and BP neural network is the most basic one. The theory introduction is shown below.
\subsubsection{The Neuron Model}
Refere to biological neural systems, the basic component of neural networks is neuron model. In a biological neural system, every neuron is connected to other neurons. A neuron accepts some chemical materials (chemical messages) as its input emanated from other connected neurons, the electrical potential in this neuron changes; If the amount of the input is enough, the electrical potential exceeds to a threshold, this neuron will be activated and send chemical materials to other neurons.

The famous McCulloch and Pitts model (M-P) of neuron as Figure 4 is based on the above description, which can be formulated in a mathematical form:
\begin{equation}
y=f(\boldsymbol{w}\boldsymbol{x}-\theta)
\end{equation}
where $\boldsymbol{x}=\{x_i\}_{i=1}^n$ is the input of the neuron, $\boldsymbol{w}=\{w_i\}_{i=1}^n$ is its corresponding weight, $\theta$ is a threshold of activation, and $f(.)$ is an activation function with multiple forms. The step function is the most ideal activation function, but it is discontinuous and not smooth, so it is not convenient to do the operations of integral and differential. Hence, in most cases, we usually use a canonical example, the sigmoid function which can squash the large range of input values into the range from 0 to 1 of outputs. The step function and the sigmoid function are shown in the Figure 5.
\begin{figure}[!htb]
	\centering\includegraphics[width=6cm]{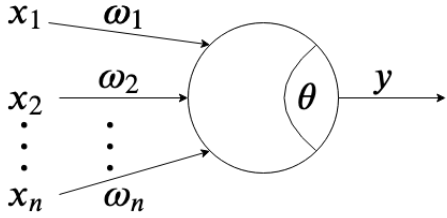}\\
	\caption{M-P Neuron Model}\label{sigmoid}
\end{figure}
\begin{figure}[!htb]
	\centering\includegraphics[width=6cm]{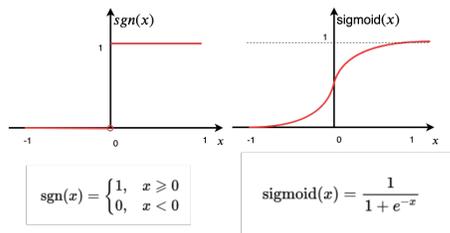}\\
	\caption{The Step Function and the Sigmoid Function}\label{sigmoid}
\end{figure}

When enough training data sets are given, the weights and threshold can be obtained by learning. The threshold can be seen as a dummy node with fixed input $1$, so the learning of weights and threshold can be jointed together. For a single training data set ($x$,$y$), the current output is $\hat{y}$ for the input $x$. Weight updating can be shown as:
\begin{equation}
w_i\leftarrow w_i+\bigtriangleup w_i
\end{equation}
\begin{equation}
\bigtriangleup w_i=\eta (y-\hat{y})x_i
\end{equation}
where $\eta \in (0,1)$ is a learning rate. In this way, the weight will be adjusted until $y = \hat{y}$. This is a single functional neuron, its learning ability is very limited. Thus, we always add the hidden layers to compose a multi-layer feedforward nerual network.

\subsubsection{Error Backpropagation}
For multi-layer networks, BP(backpropagation) is one of the most successful learning algorithms. BP algorithm is not only used in the multi-layer feed forward networks, but also used in other networks such as recurrent neural networks. The basic structure of the neural networks consists of an input layer and an output layer with multiple hidden layers between them. For clear and concise description, the number of hidden layers is set to $1$.

The input layer is $\boldsymbol{x}=\{x_i\}_{i=1}^d$, the hidden layer is $\boldsymbol{b}=\{b_h\}_{h=1}^q$, and the output layer is $\hat{\boldsymbol{y}}=\{\hat{y}\}_{j=1}^l$. The weight between the input layer node $i$ and the hidden layer node $h$ is $v_{ih}$, The weight between the hidden layer node $h$ and the output layer node $y$ is $w_{ih}$. The variable symbols and the BP network model with three layers are shown in Figure 6, the weights are also represented.
\begin{figure}[!htb]
	\centering\includegraphics[width=6cm]{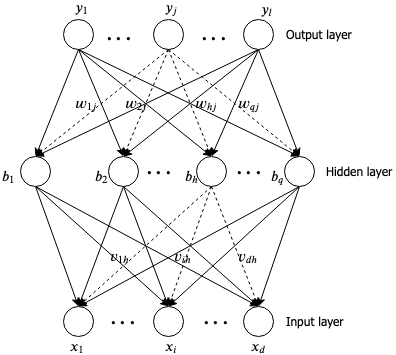}\\
	\caption{Multilayer Network Sketch}\label{MLBMS}
\end{figure}

For $h^{th}$ node of hidden layer and $j^{th}$ node of output layer, the inputs and the outputs are shown below:
\begin{equation}
    \begin{cases}
        \alpha_h=\sum_{i=1}^{d}v_i{}_h x_i\\
        b_h=f(\alpha_h-\gamma_h)\\
        \beta_j=\sum_{h=1}^{q}w_h{}_j b_h\\
        \hat{y}_j=f(\beta_j-\theta_j)
    \end{cases}
\end{equation}

Then the mean square error can be obtained as following, where $1/2$ is for the convenience of the subsequent calculation.
\begin{equation}
E=\frac{1}{2}\sum_{j=1}^{l}(\hat{y}_j-y_j)^2
\end{equation}

Because BP algorithm is based on gradient descent algorithm, parameters are adjusted according to the negative gradient. Set the gradient of output layer as an example, the change of weight can be written as
\begin{equation}
\bigtriangleup w_h{}_j=-\eta \frac{\partial E}{\partial w_h{}_j}
\end{equation}

Besides, the chain rule can make the formula as
\begin{equation}
\frac{\partial E}{\partial w_h{}_j}=\frac{\partial E}{\partial \hat{y}_j}\cdot\frac{\partial \hat{y}_j}{\partial\beta_j}\cdot\frac{\partial \beta_j}{\partial w_h{}_j}
\end{equation}
\begin{equation}
\frac{\partial\beta_j}{\partial w_h{}_j}=b_h
\end{equation}

The intermediate variable of gradient can be obtained as
\begin{equation}
g_j=-\frac{\partial E}{\partial \hat{y}_j}\cdot\frac{\partial \hat{y}_j}{\partial\beta_j}=\hat{y}_j(1-\hat{y}_j)(y_j-\hat{y}_j)
\end{equation}

Therefore, the gradient of the hidden layer is similar,
\begin{equation}
e_h=-\frac{\partial E}{\partial b_h}\cdot\frac{\partial b_h}{\partial\alpha_h}=b_h(1-b_h)\sum_{j=1}^{l}w_h{}_j
\end{equation}

Finally, we can get the update of weights and thresholds for the hidden and output layer,
\begin{equation}
\begin{cases}
   \bigtriangleup w_h{}_j=\eta g_jb_h\\
   \bigtriangleup \theta_j=-\eta g_j\\
   \bigtriangleup v_i{}_h=\eta e_hx_i\\
   \bigtriangleup \gamma_h=-\eta e_h
\end{cases}
\end{equation}

Note that the aim of BP algorithm is to minimize the accumulate error for all training data sets. The workflow of BP algorithm is represented in Algorithm 1. The stop condition is that the iteration has achieved the peak value or the error $E$ is smaller than a minimum threshold.
\begin{algorithm}
	\renewcommand{\algorithmicrequire}{\textbf{Input:}}
	\renewcommand{\algorithmicensure}{\textbf{Output:}}
	\caption{BP Algorithm \cite{6}}
	\label{alg:1}
	\begin{algorithmic}
		\REQUIRE A training data set  $\boldsymbol{S}=\{(\boldsymbol{x}_k,\boldsymbol{y}_k)\}^m_{k=1}$, a learning rate $\eta$.
		\STATE Randomly initiate all weights and thresholds $w$, $\theta$, $v$, $\gamma$
		\REPEAT
		\FORALL{$(\boldsymbol{x}_k,\boldsymbol{y}_k)\in \boldsymbol{S}$}
		\STATE Calculate the current output $\hat{y}_k$ using Equation (4.4)
		\STATE Calculate intermediate variable of gradient $g$ and $e$ \\
		using Equation (4.9) and (4.10)
		\STATE Update weights and threshold $w$, $\theta$, $v$, $\gamma$ using Equation (4.11)
		\ENDFOR
		\UNTIL{The stop condition is reached}
		\ENSURE Weights and thresholds $w$, $\theta$, $v$, $\gamma$
	\end{algorithmic}
\end{algorithm}

However, BP algorithm has some inadequacies. Overfitting problem is very obvious because of the powerful presentation of BP networks, early stopping method and regularization method are often used to prevent this situation. The limitation of local minimum is another problem, people usually utilize simulated annealing, genetic algorithm and random gradient descent algorithm to solve it \cite{6}.

\subsection{Ground User Clustering}
Ground user clustering is a key step in UAV deployement, which also means the partition of UAV aerial cells. In order to satisfy the fairness and globality of division, we adopt a KEG algorithm to implement service area classification. And To show the algorithm practicability, we use the topology information part of City Cellular Traffic Map \cite{chen2015analyzing} as the real data set.
\subsubsection{The K-means Algorithm}
The K-means algorithm is a most basic non-hierarchical iterative clustering algorithm, belonging to unsupervised learning. The goal of this algorithm is to cluster data points with very low inter-cluster similarity as well as very high intra-cluster similarity \cite{14}. The similarity often denotes the distances between data points.

Giving a data set $\boldsymbol{X}=\{\boldsymbol{x}_n\}^N_{n=1}$ composed of $N$ number of $M$-dimensional variables $\boldsymbol{x}_n$. We divide this data set into $K$ clusters, each cluster has a centroid. At first, we set the centroids of clusters are $K$ vectors $\{\boldsymbol\mu_k\}^K_{k=1}$ with $M$ dimensions. These vectors usually randomly take $K$ variables from the given data set $\boldsymbol{X}$ for initialisation. To realize the algorithm goal, each data point is supposed to be as close as possible to its cluster centroid, and the distances of each data point and other cluster centroids should be as large as possible. Thus, we set that point cluster label $\boldsymbol{r}_{nk}$ indicates which cluster the data point $\boldsymbol{x}_n$ belongs to, this variable can be formulated as:
\begin{equation}
    \boldsymbol{r}_{nk}=
    \begin{cases}
        1& \text{if $k = \mathop{\arg\min}_{i} \ \|\boldsymbol{x}_n - \boldsymbol{\mu}_i\|$}\\
        0& \text{otherwise}
    \end{cases}
\end{equation}

Then, we assign every data point to its nearest cluster and update the point cluster label after the distances between each data point and cluster centroids are calculated.  The third one is to update cluster centroids according to those data point label. The specific K-means algorithm is shown as Algorithm 2.
\begin{algorithm}
	\renewcommand{\algorithmicrequire}{\textbf{Input:}}
	\renewcommand{\algorithmicensure}{\textbf{Output:}}
	\caption{K-means Algorithm \cite{14}}
	\label{alg:K-means}
	\begin{algorithmic}[1]
		\REQUIRE The cluster number $K$, the data point set $\boldsymbol{X}=\{\boldsymbol{x}_n\}^N_{n=1}$.
		\STATE Initialize $\{\boldsymbol{\mu}_k\}^K_{k=1}$ as $K$ variables chose from $\boldsymbol{X}$ randomly, \\
	    initialize $\boldsymbol{r}_{nk}$ as an $n*k$  all-zero matrix
		\REPEAT
		\FORALL{$\boldsymbol{x}_n\in X$}
		\STATE Allocate each data point $\boldsymbol{x}_n$ to cluster
		$k^* = \mathop{\arg\min}_{i} \ \|\boldsymbol{x}_n - \boldsymbol{\mu}_i\|$
		\STATE Update point labels $\boldsymbol{r}_{nk}(n,k^*)= 1$
		\STATE Calculate cluster centroids
		$\boldsymbol{\mu}_k= \sum \boldsymbol{r}_{nk} \boldsymbol{x}_n / \sum \boldsymbol{r}_{nk}$, $k = 1,...,K$
		\ENDFOR
		\UNTIL{$\{\boldsymbol{\mu}_k\}^K_{k=1}$ will not be changed.}
		\ENSURE The point cluster label $\boldsymbol{r}_{nk}$, the cluster centroids $\{\boldsymbol{\mu}_k\}^K_{k=1}$
	\end{algorithmic}
\end{algorithm}

On the one side, although the K-means algorithm has the capability to cluster data points, it cannot find the latent variables corresponding the observed data. On the other side, the principle of the K-means is simple and it is easy to implement the simulation because of its fast convergence and good clusting effect, it can be a pratical data initializer for other complex algorithms.

\subsubsection{The EM Algorithm}
The EM algorithm has the ability to recognize the important role of latent variables in a joint distribution. Its goal is to acquire maximum likelihood results for the models with latent variables \cite{14}.

In a mixed distribution model, given a sample data set $\boldsymbol{X} = \{\boldsymbol{x}_n\}^N_{n=1}$ with an unknown latent variable set $\boldsymbol{Z} = \{\boldsymbol{z}_n\}^N_{n=1}$, we want to find the suitable parameter set $\boldsymbol{\theta}$ to well discribe the joint distribution $p(\boldsymbol{X}|\boldsymbol{\theta}) = \sum_{\boldsymbol{Z}}p(\boldsymbol{X},\boldsymbol{Z}|\boldsymbol{\theta})$ \cite{14}. However, the observed variable set $\boldsymbol{X}$ and the latent variable set $\boldsymbol{Z}$ are determined by parameter set $\boldsymbol{\theta}$. And we are only given the incomplete data set $\boldsymbol{X}$ without $\boldsymbol{Z}$, so we cannot get the optimal parameter set $\boldsymbol{\theta}$. To facilitate analysis, a log likelihood function is defined as:
\begin{equation}
    {\cal{L}}(\boldsymbol{\theta})=\ln{\{p(\boldsymbol{X}|\boldsymbol{\theta})\}} = \ln{\{\sum_{\boldsymbol{Z}}p(\boldsymbol{X},\boldsymbol{Z}|\boldsymbol{\theta})\}}
\end{equation}

And a posterior distribution $p(\boldsymbol{Z}|\boldsymbol{X},\boldsymbol{\theta})$ about the latent variable set $\boldsymbol{Z}$ is introduced. Thus, our goal is changed to get the maximum likelihood function $p(\boldsymbol{X}|\boldsymbol{\theta})$ with regard to $\boldsymbol{\theta}$.
The iteration of EM mainly consists of two step: Expectation step (E step) and Maximum step (M step). In E step, we evalulate the posterior probability $p(\boldsymbol{Z}|\boldsymbol{X},\boldsymbol{\theta)}$; In M step, we operate the log likelihood maximization to update parameter set $\boldsymbol{\theta}$ \cite{14}. The iteration stops until the convergence of the log likelihood function is checked. The specific algorithm is shown in Algorithm 3.

\begin{algorithm}
	\renewcommand{\algorithmicrequire}{\textbf{Input:}}
	\renewcommand{\algorithmicensure}{\textbf{Output:}}
	\caption{EM Algorithm \cite{14}}
	\label{alg:EM}
	\begin{algorithmic}[1]
		\REQUIRE The observed variable set $\boldsymbol{X}$
		\STATE Initialize the parameter set $\boldsymbol{\theta}_{old}$,
		\REPEAT
		\STATE \textbf{E step} Calculate the posterior probability $p(\boldsymbol{Z}|\boldsymbol{X},\boldsymbol{\theta}_{old})$
		\STATE \textbf{M step} Calculate $\boldsymbol{\theta}_{new} = \mathop{\arg\min}_{\boldsymbol{\theta}} \ \sum_{\boldsymbol{Z}} p(\boldsymbol{Z}|\boldsymbol{X},\boldsymbol{\theta}_{old}) \ln{p(\boldsymbol{X},\boldsymbol{Z}|\boldsymbol{\theta})}$
		\STATE Update the parameter set $\boldsymbol{\theta}$, $\boldsymbol{\theta}_{old} \leftarrow \boldsymbol{\theta}_{new}$
		\UNTIL{The log likelihood function ${\cal{L}}(\boldsymbol{\theta})$ converges.}
		\ENSURE The posterior probability $p(\boldsymbol{Z}|\boldsymbol{X},\boldsymbol{\theta})_{old}$, the parameter set $\boldsymbol{\theta}_{old}$
	\end{algorithmic}
\end{algorithm}

The EM algorithm can also have good performance in the situation of missing some observed variable values, the observed variable distribution can be aquired by marginializing those missing value and taking the whole joint variable distribution. In this case, the data returned by the sensor which has some values missing can also be well processed. Therefore, in the scenario of our ground user clustering, EM algorithm is a useful method to find the latent variables in data set and classify users in a fair manner.
\subsubsection{The KEG Algorithm}
In this paper, the cellular traffic distribution is complex and time-varying, but a GMM, a linear Gaussian-component superposition model has a remarkable advantage of abundantly representing data distribution. We model the cellular traffic distribution by the GMM as:
\begin{equation}
    p(\boldsymbol{X})=\sum_{k=1}^K \pi_k {\cal{N}}(\boldsymbol{X}|\boldsymbol{\mu}_k,\boldsymbol{\sigma}_k)
\end{equation}
where $\boldsymbol{X}=\{\boldsymbol{x}_n\}^N_{n = 1}$ is the tolopogy information of the whole area and $\boldsymbol{x}_n$ is every data point with $M$ dimensions. $p(.)$ represents as a probability function, $K$ is Gaussian component number and $ k \in \{1,...,K\}$ denotes a specific one Gaussian component, the mixing coefficient $\pi$ equals one or zero has $\sum_{k=1}^K \pi_k = 1$, $\boldsymbol{\mu}=\{\boldsymbol{\mu}_n\}^N_{n = 1}$ is the mean values corresponding to the cluster centroids with $M$ dimensions, $\boldsymbol{\sigma}=\{\boldsymbol{\sigma}_n\}^N_{n = 1}$ is the covariance with $M$ dimensions. Besides, in a GMM, the latent variables are discrete.

To this end, we introduce the KEG algorithm based on the K-means algorithm and the EM algorithm.
In the KEG algorithm, the EM part aims to find the discrete latent variables and the suitable parameters for analyzing data distribution and clustering data set. Even if a data set is incomplete with missing some values, the EM can also process the data set in a suitable manner. In the EM part, the value of the log likelihood function will increase with the number of iteration rising. When the log likelihood function does not change anymore, the current parameters are the aims we want to obtain. But since this algorithm usually needs many iterations to reach the convergent point, the K-means part is utilized as a data initializer to provide appropriate and rational initialized values. The integrated KEG algorithm is demonstrated in Algorithm 4.
\begin{algorithm}
	\caption{KEG algorithm}
	\label{alg:KEG}
	\renewcommand{\algorithmicrequire}{\textbf{Input:}}
	\renewcommand{\algorithmicensure}{\textbf{Output:}}
	\begin{algorithmic}[1]
		\REQUIRE The topology data set $\boldsymbol{X} = \{\boldsymbol{x}_n\}^N_{n = 1}$, the clustering number $K$
		\STATE Set a variable $D$ as the dimension of $\boldsymbol{x}_n$, initialize the means $\boldsymbol{\mu}_k$ as a $1$-by-$D$ matrix by using the K-means algorithm, the covariance $\boldsymbol{\sigma}_k$ as a $D$-by-$D$ identical matrix and mixing coefficients $\pi_k = 1/K$, $\forall k \in \{1,...,K\}$.
		\REPEAT
		\FORALL{$k \in \{1,...,K\}$}
		\STATE \textbf{E step:} Compute the posterior probability of all $\boldsymbol{x_n}$ by\\
		$\gamma_{nk} =\pi_k {\cal{N}}(\boldsymbol{x}_n|\boldsymbol{\mu}_k,\boldsymbol{\sigma}_k) / \sum_{i=1}^K \pi_i{\cal{N}}(\boldsymbol{x}_n|\boldsymbol{\mu}_i,\boldsymbol{\sigma}_i)$
		\STATE For every $\boldsymbol{x}_n$, allocate it to the cluster $l_n = \mathop{\arg\max}_{k} \gamma_{nk} $
		\STATE \textbf{M step:} Calculate the new parameters \\
		$\boldsymbol{\mu}_k^{new} = \sum_{n=1}^N \gamma_{nk}\boldsymbol{x}_n/\sum_{n=1}^N \gamma_{nk}$ \\
		$\boldsymbol{\sigma}_k^{new} = \sum_{n=1}^N \gamma_{nk}(\boldsymbol{x}_n-\boldsymbol{\mu}_k^{new})(\boldsymbol{x}_n-\boldsymbol{\mu}_k^{new})^T/\sum_{n=1}^N \gamma_{nk}$\\
		$\pi_k^{new} = \sum_{n=1}^N \gamma_{nk} / \sum_{k=1}^K\sum_{n=1}^N \gamma_{nk}$
		\ENDFOR
		\STATE Calculate the log likelihood using $\boldsymbol{\mu}_k^{new}$, $\boldsymbol{\sigma}_k^{new}$ and $\pi_k^{new}$, for $k \in \{1,...,K\}$\\
		${\cal{L}}(\boldsymbol{\mu},\boldsymbol{\sigma},\boldsymbol{\pi})=\ln{p(\boldsymbol{X}|\boldsymbol{\mu},\boldsymbol{\sigma},\boldsymbol{\pi})} = \sum_{n=1}^N \ln{\{\sum_{k=1}^K \pi_k {\cal{N}}(\boldsymbol{x}_n|\boldsymbol{\mu}_k,\boldsymbol{\sigma}_k)\}}$
		\STATE Update the parameters as
		$\boldsymbol{\mu}_k \leftarrow \boldsymbol{\mu}_k^{new}$, $\boldsymbol{\sigma}_k \leftarrow \boldsymbol{\sigma}_k^{new}$, $\pi_k \leftarrow \pi_k^{new}$
		\UNTIL{The log likelihood function ${\cal{L}}(\boldsymbol{\mu},\boldsymbol{\sigma},\boldsymbol{\pi})$ is converged.}
		\ENSURE The parameters $\{\pi_k$, $\boldsymbol{\mu}_k$, $\boldsymbol{\sigma}_k\}_{k=1}^K$, the cluster labels $\{l_n\}_{n=1}^N$.
	\end{algorithmic}
\end{algorithm}

When the parameters are obtained, the predicted cellular traffic amount data is used as the input, only $4^{th}$ step and $5^{th}$ step of Algorithm 4 is operated, we can get the cluster label $\{l_n\}_{n=1}^N$of the data points $\{\boldsymbol{x}_n\}^N_{n = 1}$ that indicates which cluster the data point belongs to.

\subsection{UAV Location Optimization}
After determining the aerial cells using the KEG algorithm, our next aim is to select a optimal location for every UAV so that the minimum transmit power can be gotten. No matter whether UAVs are in a high altitude platform or a low altitude platform, we assume that all UAVs are in the same altitude $h$, we can formulate this problem as:
\begin{equation}
    \min_{x_i, y_i} \ \ P_i = Q \iint_{C_i} A_i(x,y)d_i^2(x,y)r_i(x,y)dxdy
\end{equation}
where $Q = (\frac{4\pi f_c}{c})^2\frac{Wn_0}{G}$ is not relevant to the locations of BS $(x,y)$ in the service area, $A_i(x,y) = 2^{\frac{TD^t(x,y)}{W}-1}$ is the BS distribution and $D^t(x,y)$ denotes the cellular data amount of BS at $(x,y)$ in a hour in the predictive day, $d_i^2(x,y)$ is the distance between the UAV $i$ and the BS at $(x,y)$, $r_i(x,y)$ is also related to $d_i^2(x,y)$ for the computation of LoS link probability.

According to the Theorem 1 of paper \cite{7}, the function to get $P_i^{min}$ is a convex function, UAV optimal locations can be calculated as:
\begin{equation}
    x_i^* = \frac{\iint_{C_i}xA_i(x,y)dxdy}{\iint_{C_i}A_i(x,y)dxdy}
\end{equation}
\begin{equation}
    y_i^* = \frac{\iint_{C_i}yA_i(x,y)dxdy}{\iint_{C_i}A_i(x,y)dxdy}
\end{equation}
where there is a condition $h_i>>(x-x_i)^2 + (y-y_i)^2$ or $h_i<<(x-x_i)^2 + (y-y_i)^2$ to be satisfied \cite{7}.

At last, we accumulate the transmit power of all operating UAV to get the minimum total power for transmission:
\begin{equation}
    P_{min} = \sum_{i} P_i^{min}
\end{equation}

\section{Simulation and Analysis}
\subsection{Illustration of Simulation Process}
For simulation, we consider two scenrios, one is the entire area given from the raw data, the other is the limited area with the relative longitude from $111.055$ to $111.07$ degrees and the relative altitude from $13.03$ to $13.05$ degrees. The limited area is the contrast of the entire area. And for a specific contrast, we classify the entire area and the limited area into $8$ clusters. Some parameters used in the simulation have been given the specific values which are shown in Table 1 \cite{yang2020energyefficient}. Moreover, for getting the rational simulation results, we assume that all BSs have a basic celluar amount $500$ bytes for basic operation, so all the cellular amounts we get from the raw data add $500$.

\begin{table}
\centering
\caption{Simulation Parameters}\label{table}
\begin{tabular}{@{}ccc@{}}
$f_c$                         & Carrier frequency                     & 5GHz       \\
$n_o$                         & Noise power spectral                  & -140dBm/Hz \\
$\mu_{LoS}$  & Excessive path loss for LoS link           & 3dB        \\
$\mu_{NLoS}$ & Excessive path loss for NLoS link         & 23dB       \\
$W$                            & Bandwidth                             & 1MHz       \\
$h$                            & UAV's altitude                        & 200m       \\
$G$                            & Antenna gain                          & 10dB       \\
\end{tabular}
\end{table}

For BP neural network part, we use the \emph{nftool} built-in box of MATLAB to implement the cellular traffic amount prediction.
We assume that every BS has the same cellular traffic distribution, we set that the input is the cellular traffic amount of the first six days and the output is the amount of the seventh day. The input in the training data set form is $(\{\boldsymbol{D}^t_d\}_{d =1}^5, \boldsymbol{D}^t_{d=6})$, where the input layer is the first $5$ days cellular amount and the output layer is the $6^{th}$ day cellular amount.

We choose \emph{trainrp} as the training function and set that the neural network has 2 hidden layers with $20$ and $10$ neurons respectively, the network structure is shown in Figure 7. Then we do data training to find the suitable parameters for our neural network model. The user interface of neural network training is shown in Figure 8.
\begin{figure}[!htb]
	\centering\includegraphics[width=6cm]{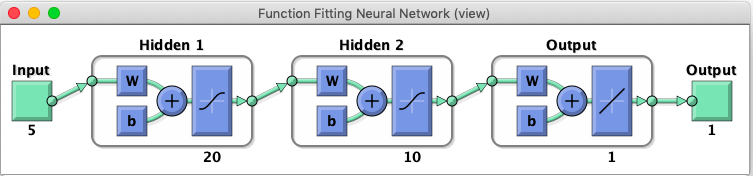}\\
	\caption{Neural Network Structure}\label{nftool}
\end{figure}
\begin{figure}[!htb]
	\centering\includegraphics[width=6cm]{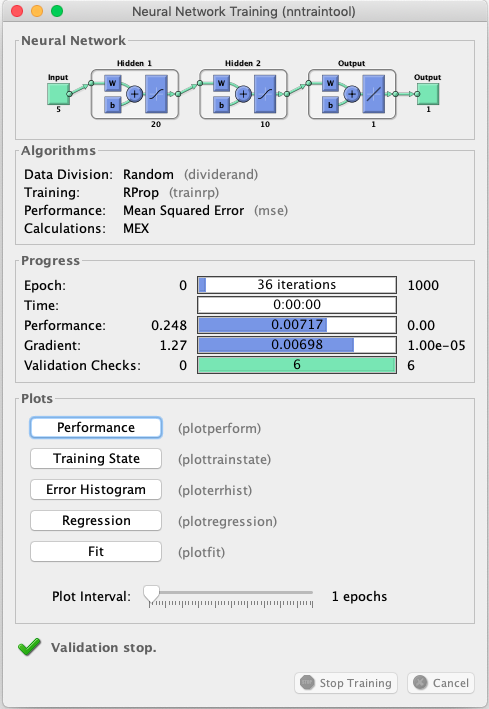}\\
	\caption{Neural Network Training}\label{nnt}
\end{figure}

After training, the input layer $\boldsymbol{x}$ is set as $\{\boldsymbol{D}^t_d\}_{d =2}^6$, then this output layer is what we want to obtain. Furthermore, the ratio of training data is $80\%$, valuating data is $10\%$ and testing data is $10\%$. The data training is for updating weights, the data evaluation is to detect the overfitting problem and to avoid it as soon as possible, the data testing is to examine the performance of the neuron network.
In addition, all the cellular data amounts are normalized using min-normalizaion mothed to make sure every single value is ranged from $0$ to $1$.

Then, for the application of the KEG algorithm, we give two conditions for finishing the iterations: the first one is the minimum threshold for parameter error of two neighboring iterations, the second one is the maximum iteration times. If any of these conditions are met, the algorithm ends.

\subsection{The Comparision of Three Multi-access Techniques}
In the proposed UAV deployment framework, RSMA as an excellent option of multiple access technique is adopted in the uplink transmission. FDMA and TDMA have been developed maturely, but RSMA with excellent robustness and energy effeciency enjoys great popularity in the new generation of communications.

The Figure 9 is shown below which represents the comparison the performance of RSMA, FDMA and TDMA.
For all three of multi-access techniques, with the increasing of bandwidth, the system can reduce the total power. But RSMA is the best one which has the lowest power consumption in all bandwidths. Then FDMA is the second best and it is better than TDMA. RSMA can reduce 35.3\% power than FDMA and 66.4\% than TDMA. Therefore, in this paper, we use RSMA for uplink transmission.
\begin{figure}[!htb]
	\centering\includegraphics[width=6cm]{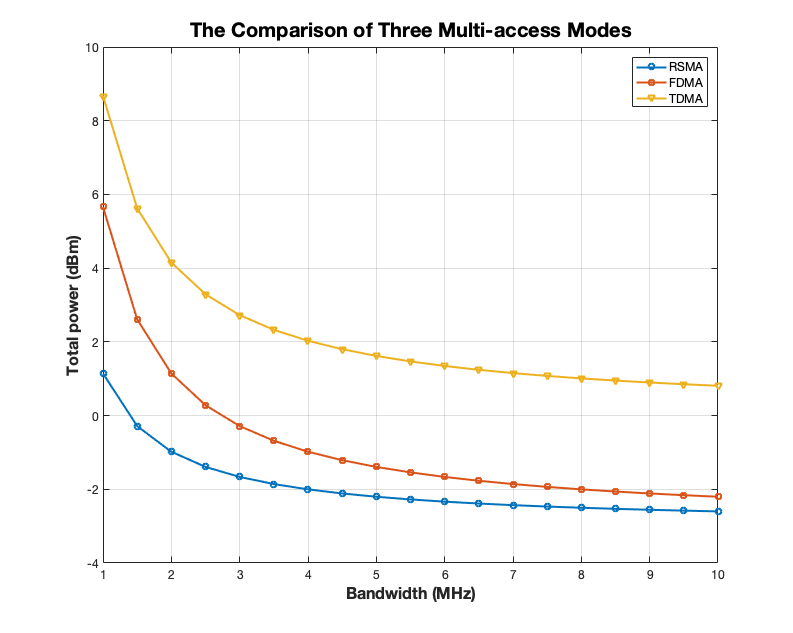}\\
	\caption{Comparison of Three Multi-access Modes}\label{RSMA}
\end{figure}

\subsection{The Simulation Results of the Limited Area and the Entire Area}
In order to verify the usefulness and robustness of the KEG algorithm, we use the data of the limited area and the entire area to do the simulation. The comparison results of the K-means part are shown in Figure 10 and Figure 11 and the EM part are shown in Figure 12 and Figure 13.


\begin{figure}[!htb]
	\centering\includegraphics[width=6cm]{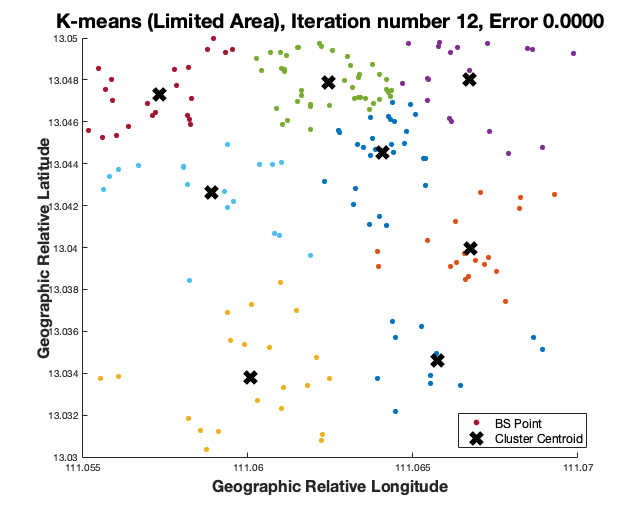}\\
	\caption{K-means for the Limited Area}\label{kmeans}
\end{figure}
\begin{figure}[!htb]
	\centering\includegraphics[width=6cm]{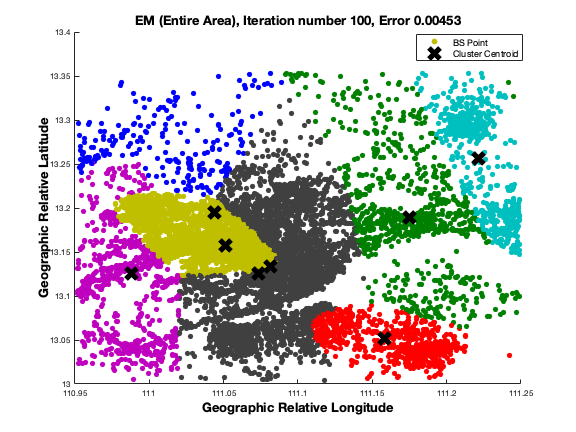}\\
	\caption{EM for the Entire Area}\label{kmeans}
\end{figure}
\begin{figure}[!htb]
	\centering\includegraphics[width=6cm]{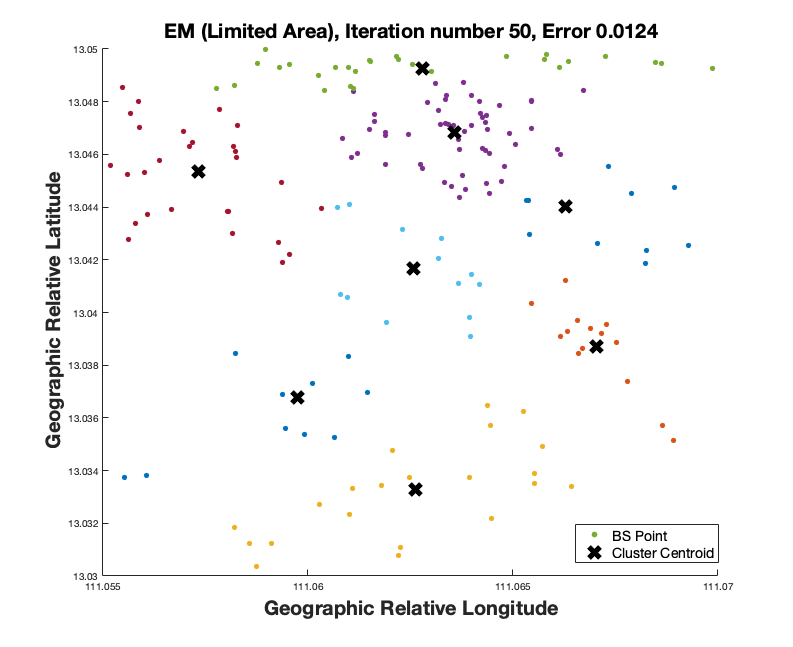}\\
	\caption{EM for the Limited Area}\label{kmeans}
\end{figure}

In the Figure 9, 10, 11 and 12, the scattering points are the locations of BS. The group of the points with the same color represents a cluster which is an aerial cell of a UAV. The black crosses denotes the centroids of every cluster. As we can see, for the K-means part, both the entire area and the limited area have distinct boundaries of service cell and the centroids locate at the clusters with the corresponding colors. But for the EM part, the clusters have the containing and contained states, the centroids may not appear in the clusters with their own colors. The corresponding situation of the limited area is much better than that the entire area.

\subsection{The Effect of Proposed UAV Scheduling Framework}
The main goal of UAV deployment in this paper is to minimize the total transmit power in the uplink transmission. Thus, based on this goal, we divide the area into the ground user clusters for aerial cell classification and determine the optimal locations of UAVs. The Figure 14 shows the schematic sketch of UAV deployment for the entire area and the limited area.
\begin{figure}[!htb]
	\centering\includegraphics[width=6cm]{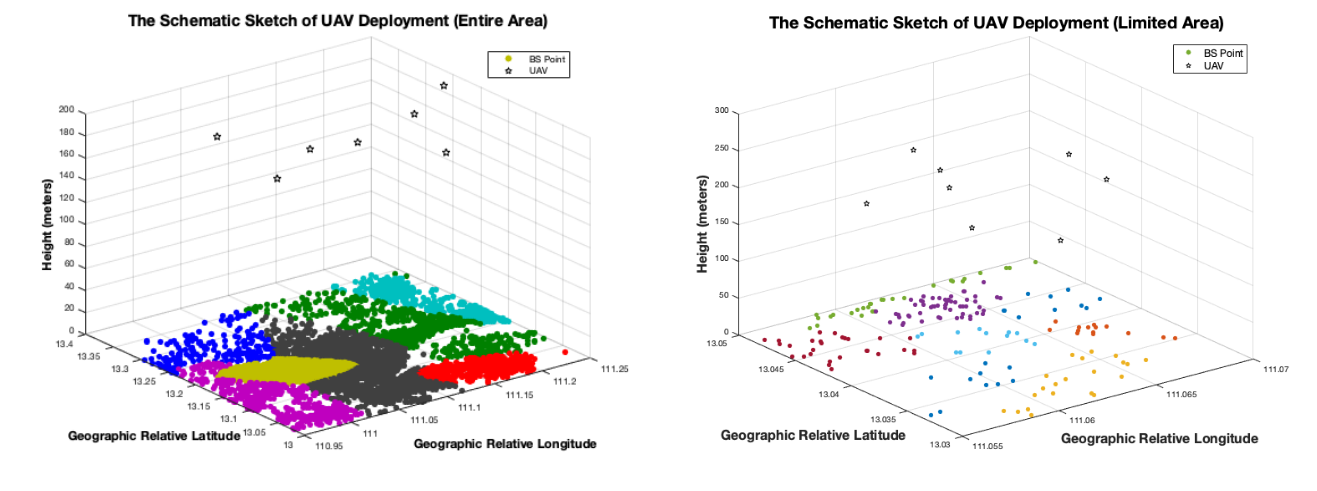}\\
	\caption{Deployment Results}\label{DR}
\end{figure}
As shown in the Figure 14, the number of UAV of the limited area is $8$, but the one of the entire area changes into $7$ because the system with our proposed framework judges that $7$ UAVs are enough to supply the offloaded cellular data amount, this decision is made during the clustering of the K-means part. Because the initial centroid of the merged category chooses an unfavorable position leading to a low similarity in the cluster.
\begin{figure}[!htb]
	\centering\includegraphics[width=6cm]{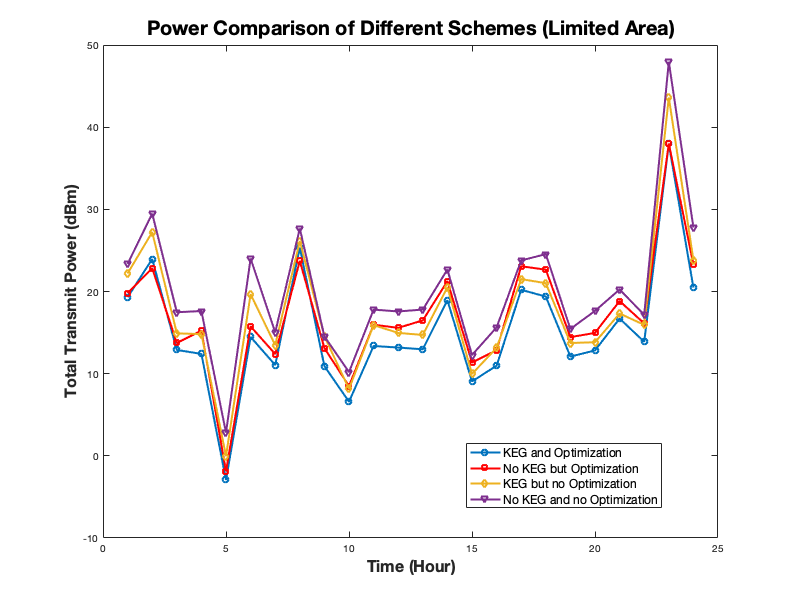}\\
	\caption{Power Comparison for the Limited Area}\label{PCLA}
\end{figure}
\begin{figure}[!htb]
	\centering\includegraphics[width=6cm]{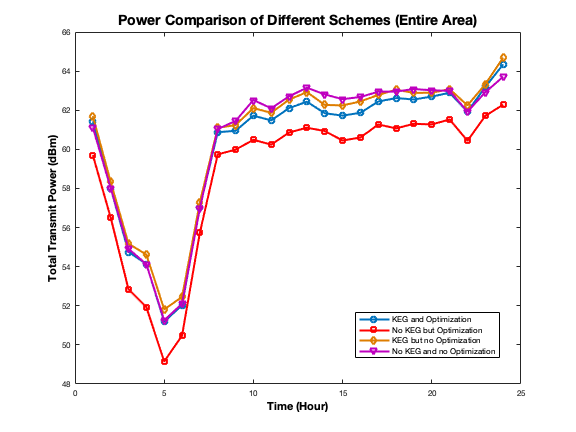}\\
	\caption{Power Comparison for the Entire Area}\label{PCEA}
\end{figure}

Then we use the total transmit power of four different schemes to evaluate the performance of the proposed framework.
Experimental schemes are a scheme with KEG and location optimization, a scheme without KEG but with location optimization, a scheme with KEG but without location optimization, and a scheme without neither KEG or location optimization.

For the limited area, the power comparison of four schemes is represented in Figure 15. In general, the scheme with KEG and location optimization consume the least total transmit power. The performance of the one with no KEG and no location optimization is worst. The remaining two types of scheme have similar performance in general. The scheme with KEF and location optimization reduce 24\% power consumption than the worst one.
For the entire area, four schemes have been contrasted in Figure 16. Obviously, the scheme without KEG but with location optimization has the best effect for system performance improvement. Next, the scheme with KEG but no location optimization is better than the one with location optimization but no KEG. The scheme with no KEG and no location is the worst scheme. The scheme with KEF and location optimization reduce 0.47\% power consumption than the worst one.

Based on the above simulation results, our proposed framework is not suitable to apply only several UAVs in the entire area. The number of UAVs is too small to carry the full range of traffic of an entire city. But still in this situation, the scheme with KEG and location optimization has some contributions for reducing the total transmit power consumption. Our UAV deployment framework has good performance for relatively small area below dozens of square kilometers, especially for those areas with the approximate GMM cellular data amount distribution.

\section{Conclusion}
In this paper, we have investigated the UAV location optimization in a downlink system.
To effectively optimize the UAV location, we fist predice the user traffic distribution by a joint K-means and EM algorithm based on GMM.
With the predicted traffic distribution, the optimal locations for UAVs are accordingly obtained.
Simulation results show that RSMA reduces total power consumption compared to FDMA, and TDMA.

\bibliographystyle{IEEEtran}
\bibliography{IEEEabrv,MMM}

\begin{thebibliography}{10}
\providecommand{\url}[1]{#1}
\csname url@samestyle\endcsname
\providecommand{\newblock}{\relax}
\providecommand{\bibinfo}[2]{#2}
\providecommand{\BIBentrySTDinterwordspacing}{\spaceskip=0pt\relax}
\providecommand{\BIBentryALTinterwordstretchfactor}{4}
\providecommand{\BIBentryALTinterwordspacing}{\spaceskip=\fontdimen2\font plus
\BIBentryALTinterwordstretchfactor\fontdimen3\font minus
  \fontdimen4\font\relax}
\providecommand{\BIBforeignlanguage}[2]{{%
\expandafter\ifx\csname l@#1\endcsname\relax
\typeout{** WARNING: IEEEtran.bst: No hyphenation pattern has been}%
\typeout{** loaded for the language `#1'. Using the pattern for}%
\typeout{** the default language instead.}%
\else
\language=\csname l@#1\endcsname
\fi
#2}}
\providecommand{\BIBdecl}{\relax}
\BIBdecl

\bibitem{saad2019vision}
W.~Saad, M.~Bennis, and M.~Chen, ``A vision of {6G} wireless systems:
  {A}pplications, trends, technologies, and open research problems,''
  \emph{arXiv preprint arXiv:1902.10265}, 2019.

\bibitem{2}
J.~Lyu, Y.~Zeng, and R.~Zhang, ``Uav-aided offloading for cellular hotspot,''
  vol.~17, no.~6.\hskip 1em plus 0.5em minus 0.4em\relax IEEE, 2018, pp.
  3988--4001.

\bibitem{chen2020wireless}
M.~Chen, H.~V. Poor, W.~Saad, and S.~Cui, ``Wireless communications for
  collaborative federated learning in the internet of things,'' \emph{arXiv
  preprint arXiv:2006.02499}, 2020.

\bibitem{3}
M.~Mozaffari, W.~Saad, M.~Bennis, Y.-H. Nam, and M.~Debbah, ``A tutorial on
  uavs for wireless networks: Applications, challenges, and open
  problems.''\hskip 1em plus 0.5em minus 0.4em\relax IEEE, 2019.

\bibitem{20}
M.~Mamdouh, M.~A. Elrukhsi, and A.~Khattab, ``Securing the internet of things
  and wireless sensor networks via machine learning: A survey,'' in \emph{2018
  International Conference on Computer and Applications (ICCA)}.\hskip 1em plus
  0.5em minus 0.4em\relax IEEE, 2018, pp. 215--218.

\bibitem{4}
J.~Zhang, X.~Zhu, and Z.~Zhou, ``Design of time delayed control systems in uav
  using model based predictive algorithm,'' in \emph{2010 2nd International
  Asia Conference on Informatics in Control, Automation and Robotics (CAR
  2010)}, vol.~1.\hskip 1em plus 0.5em minus 0.4em\relax IEEE, 2010, pp.
  269--272.

\bibitem{8637952}
Z.~{Li}, M.~{Chen}, C.~{Pan}, N.~{Huang}, Z.~{Yang}, and A.~{Nallanathan},
  ``Joint trajectory and communication design for secure uav networks,''
  \emph{IEEE Commun. Lett.}, vol.~23, no.~4, pp. 636--639, April 2019.

\bibitem{8379427}
Z.~{Yang}, C.~{Pan}, M.~{Shikh-Bahaei}, W.~{Xu}, M.~{Chen}, M.~{Elkashlan}, and
  A.~{Nallanathan}, ``Joint altitude, beamwidth, location, and bandwidth
  optimization for {UAV}-enabled communications,'' \emph{IEEE Commun. Lett.},
  vol.~22, no.~8, pp. 1716--1719, Aug. 2018.

\bibitem{8755300}
M.~{Chen}, U.~{Challita}, W.~{Saad}, C.~{Yin}, and M.~{Debbah}, ``Artificial
  neural networks-based machine learning for wireless networks: A tutorial,''
  \emph{IEEE Commun. Surveys Tut.}, vol.~21, no.~4, pp. 3039--3071,
  Fourthquarter 2019.

\bibitem{dong2019deep}
P.~Dong, H.~Zhang, G.~Y. Li, I.~S. Gaspar, and N.~NaderiAlizadeh, ``Deep
  cnn-based channel estimation for mmwave massive mimo systems,'' \emph{IEEE J.
  Sel. Topics Signal Process.}, vol.~13, no.~5, pp. 989--1000, 2019.

\bibitem{shi2020communication}
Y.~Shi, K.~Yang, T.~Jiang, J.~Zhang, and K.~B. Letaief,
  ``Communication-efficient edge ai: Algorithms and systems,'' \emph{arXiv
  preprint arXiv:2002.09668}, 2020.

\bibitem{jia2020channel}
G.~Jia, Z.~Yang, H.-K. Lam, J.~Shi, and M.~Shikh-Bahaei, ``Channel assignment
  in uplink wireless communication using machine learning approach,''
  \emph{arXiv preprint arXiv:2001.03952}, 2020.

\bibitem{chen2019joint}
M.~Chen, Z.~Yang, W.~Saad, C.~Yin, H.~V. Poor, and S.~Cui, ``A joint learning
  and communications framework for federated learning over wireless networks,''
  \emph{arXiv preprint arXiv:1909.07972}, 2019.

\bibitem{yang2019energy}
Z.~Yang, M.~Chen, W.~Saad, C.~S. Hong, and M.~Shikh-Bahaei, ``Energy efficient
  federated learning over wireless communication networks,'' \emph{arXiv
  preprint arXiv:1911.02417}, 2019.

\bibitem{wang2019caching}
Y.~Wang and V.~Friderikos, ``Caching as an image characterization problem using
  deep convolutional neural networks,'' \emph{arXiv preprint arXiv:1907.07263},
  2019.

\bibitem{8}
M.~Mozaffari, W.~Saad, M.~Bennis, and M.~Debbah, ``Efficient deployment of
  multiple unmanned aerial vehicles for optimal wireless coverage,'' vol.~20,
  no.~8.\hskip 1em plus 0.5em minus 0.4em\relax IEEE, 2016, pp. 1647--1650.

\bibitem{7}
------, ``Optimal transport theory for power-efficient deployment of unmanned
  aerial vehicles,'' in \emph{2016 IEEE international conference on
  communications (ICC)}.\hskip 1em plus 0.5em minus 0.4em\relax IEEE, 2016, pp.
  1--6.

\bibitem{wang2018energy}
L.~Wang and S.~Zhou, ``Energy-efficient uav deployment with flexible functional
  split selection,'' in \emph{2018 IEEE 19th International Workshop on Signal
  Processing Advances in Wireless Communications (SPAWC)}.\hskip 1em plus 0.5em
  minus 0.4em\relax IEEE, 2018, pp. 1--5.

\bibitem{wang2019adaptive}
Z.~Wang, L.~Duan, and R.~Zhang, ``Adaptive deployment for uav-aided
  communication networks,'' \emph{IEEE Transactions on Wireless
  Communications}, vol.~18, no.~9, pp. 4531--4543, 2019.

\bibitem{zhang2018predictive}
Q.~Zhang, W.~Saad, M.~Bennis, X.~Lu, M.~Debbah, and W.~Zuo, ``Predictive
  deployment of uav base stations in wireless networks: Machine learning meets
  contract theory,'' \emph{arXiv preprint arXiv:1811.01149}, 2018.

\bibitem{wang2019deep}
Y.~Wang, M.~Chen, Z.~Yang, T.~Luo, and W.~Saad, ``Deep learning for optimal
  deployment of uavs with visible light communications,'' \emph{arXiv preprint
  arXiv:1912.00752}, 2019.

\bibitem{7875131}
M.~{Chen}, M.~{Mozaffari}, W.~{Saad}, C.~{Yin}, M.~{Debbah}, and C.~S. {Hong},
  ``Caching in the sky: Proactive deployment of cache-enabled unmanned aerial
  vehicles for optimized quality-of-experience,'' \emph{IEEE J. Sel. Areas
  Commun.}, vol.~35, no.~5, pp. 1046--1061, May 2017.

\bibitem{11}
S.~Q. Zhang, F.~Xue, N.~A. Himayat, S.~Talwar, and H.~Kung, ``A machine
  learning assisted cell selection method for drones in cellular networks,'' in
  \emph{2018 IEEE 19th International Workshop on Signal Processing Advances in
  Wireless Communications (SPAWC)}.\hskip 1em plus 0.5em minus 0.4em\relax
  IEEE, 2018, pp. 1--5.

\bibitem{1}
Q.~Zhang, M.~Mozaffari, W.~Saad, M.~Bennis, and M.~Debbah, ``Machine learning
  for predictive on-demand deployment of uavs for wireless communications,'' in
  \emph{2018 IEEE Global Communications Conference (GLOBECOM)}.\hskip 1em plus
  0.5em minus 0.4em\relax IEEE, 2018, pp. 1--6.

\bibitem{36}
A.~Al-Hourani, S.~Kandeepan, and S.~Lardner, ``Optimal lap altitude for maximum
  coverage,'' \emph{IEEE Wireless Communications Letters}, vol.~3, no.~6, pp.
  569--572, 2014.

\bibitem{45}
Z.~Yang, M.~Chen, W.~Saad, W.~Xu, and M.~Shikh-Bahaei, ``Sum-rate maximization
  of uplink rate splitting multiple access ({RSMA}) communication,''
  \emph{arXiv preprint arXiv:1906.04092}, 2019.

\bibitem{yang2019optimization}
Z.~Yang, M.~Chen, W.~Saad, and M.~Shikh-Bahaei, ``Optimization of rate
  allocation and power control for rate splitting multiple access ({RSMA}),''
  \emph{arXiv preprint arXiv:1903.08068}, 2019.

\bibitem{7470942}
B.~{Clerckx}, H.~{Joudeh}, C.~{Hao}, M.~{Dai}, and B.~{Rassouli}, ``Rate
  splitting for {MIMO} wireless networks: {A} promising {PHY}-layer strategy
  for {LTE} evolution,'' \emph{IEEE Commun. Mag.}, vol.~54, no.~5, pp. 98--105,
  May 2016.

\bibitem{mao2018rate}
Y.~Mao, B.~Clerckx, and V.~O.~K. Li, ``Rate-splitting multiple access for
  downlink communication systems: {B}ridging, generalizing, and outperforming
  {SDMA} and {NOMA},'' \emph{EURASIP J. Wireless Commun. Network.}, vol. 2018,
  no.~1, pp. 1--54, May 2018.

\bibitem{mao2019rate}
------, ``Rate-splitting for multi-user multi-antenna wireless information and
  power transfer,'' \emph{arXiv preprint arXiv:1902.07851}, 2019.

\bibitem{8846706}
Y.~{Mao}, B.~{Clerckx}, and V.~O.~K. {Li}, ``Rate-splitting for multi-antenna
  non-orthogonal unicast and multicast transmission: {S}pectral and energy
  efficiency analysis,'' \emph{IEEE Trans. Commun.}, to appear, 2019.

\bibitem{8491100}
------, ``Energy efficiency of rate-splitting multiple access, and performance
  benefits over {SDMA} and {NOMA},'' in \emph{Proc. IEEE Int. Symp. Wireless
  Commun. Sys.}, Lisbon, Portugal, Aug. 2018, pp. 1--5.

\bibitem{8008852}
M.~{Dai} and B.~{Clerckx}, ``Multiuser millimeter wave beamforming strategies
  with quantized and statistical {CSIT},'' \emph{IEEE Trans. Wireless Commun.},
  vol.~16, no.~11, pp. 7025--7038, Nov. 2017.

\bibitem{clerckx2019rate}
B.~Clerckx, Y.~Mao, R.~Schober, and H.~V. Poor, ``Rate-splitting unifying
  {SDMA}, {OMA}, {NOMA}, and multicasting in {MISO} broadcast channel: {A}
  simple two-user rate analysis,'' \emph{arXiv preprint arXiv:1906.04474},
  2019.

\bibitem{chen2015analyzing}
X.~Chen, Y.~Jin, S.~Qiang, W.~Hu, and K.~Jiang, ``Analyzing and modeling
  spatio-temporal dependence of cellular traffic at city scale,'' in \emph{2015
  IEEE International Conference on Communications (ICC)}.\hskip 1em plus 0.5em
  minus 0.4em\relax IEEE, 2015, pp. 3585--3591.

\bibitem{44}
U.~Paul, A.~P. Subramanian, M.~M. Buddhikot, and S.~R. Das, ``Understanding
  traffic dynamics in cellular data networks,'' in \emph{2011 Proceedings IEEE
  INFOCOM}.\hskip 1em plus 0.5em minus 0.4em\relax IEEE, 2011, pp. 882--890.

\bibitem{6}
Z.~Zhou, ``Machine learning,'' in \emph{Bioinformatics (Oxford, England)},
  2010.

\bibitem{14}
C.~M. Bishop, ``Pattern recognition and machine learning.''\hskip 1em plus
  0.5em minus 0.4em\relax springer, 2006.

\bibitem{yang2020energyefficient}
Z.~Yang, M.~Chen, W.~Saad, W.~Xu, M.~Shikh-Bahaei, H.~V. Poor, and S.~Cui,
  ``Energy-efficient wireless communications with distributed reconfigurable
  intelligent surfaces,'' 2020.

\end{thebibliography}

\end{document}